\documentclass{PoS}

\usepackage{graphicx}
\usepackage{grffile}
\usepackage{amsmath}
\usepackage[utf8]{inputenc}
\usepackage{psfrag}

\usepackage{wrapfig}

\title{Simulations of $\mathcal{N} = 1$ supersymmetric Yang-Mills theory with three colours}

\ShortTitle{Simulations of $\mathcal{N} = 1$ supersymmetric Yang-Mills theory with three colours}

\author{Sajid Ali, Henning Gerber, \speaker{Pietro Giudice}, Gernot M\"unster\\
  Universit\"at M\"unster, Institut f\"ur Theoretische Physik, \\
  Wilhelm-Klemm-Str. 9, D-48149 M\"unster, Germany\\
  E-mail: \email{sajid.ali@uni-muenster.de, p.giudice@uni-muenster.de, h.gerber@uni-muenster.de, munsteg@uni-muenster.de}}

\author{Georg Bergner\\
  Universit\"at Bern, Institut f\"ur Theoretische Physik, \\
  Sidlerstr.~5, CH-3012 Bern, Switzerland\\
  E-mail: \email{bergner@itp.unibe.ch}
}

\author{Istvan Montvay\\
  Deutsches Elektronen-Synchrotron DESY, \\
  Notkestr. 85, D-22603 Hamburg, Germany\\
  E-mail: \email{montvay@mail.desy.de}
}

\author{Stefano Piemonte\\
Universität Regensburg, Institute for Theoretical Physics, \\
D-93040 Regensburg, Germany \\
  E-mail: \email{stefano.piemonte@ur.de}
}

\abstract{
We report on our recent results regarding numerical simulations of the four 
dimensional, $\mathcal{N} = 1$ Supersymmetric Yang-Mills theory with 
SU(3) gauge symmetry and light dynamical gluinos.
}

\FullConference{34th annual International Symposium on Lattice Field Theory\\
         24-30 July 2016\\
         University of Southampton, UK}

\newcommand{\aetap}{\text{a--}\eta'}
\newcommand{\api}{{\text{a--}\pi}}
\newcommand{\afn}{\text{a--}f_0}

\newcommand{\tr}[1]{\ensuremath{\mathrm{Tr}} \left[{#1}\right]}
\newcommand{\I}{\ensuremath{\mathrm{i}\hspace{1pt}}}

\newcommand{\beq}{\begin{equation}}
\newcommand{\eeq}{\end{equation}}
\newcommand{\bea}{\begin{eqnarray}}
\newcommand{\eea}{\end{eqnarray}}

\begin{document}
\section{Introduction}

In this work we explore the supersymmetric Yang-Mills theory (SYM) with 
gauge group SU(3). This work is a natural continuation of what has been done
by our collaboration until now simulating the gauge group SU(2).
Our conclusive results have been presented in~\cite{Bergner:2015adz}:
we have verified that, in the continuum limit, the degeneracy of the 
supermultiplet is recovered and we have no sign of a possible
spontaneous breaking of supersymmetry (SUSY). 
Moreover, in~\cite{Bergner:2014saa} we have studied the theory at finite 
temperature: the most interesting result was the evidence that
chiral symmetry is restored near the deconfinement phase transition.

SU(2) SYM has been an interesting test case for the more phenomenological 
relevant SU(3) theory, which contains the gluons of QCD. 
On the other hand, there are important new aspects in SU(3) SYM like the 
new bound states and CP-violating phases. From the computational side SU(3) 
is much more demanding than QCD and SU(2) SYM. 
Moreover, this model has been proposed as an attractive candidate for a 
supersymmetric hidden dark-matter sector that may explain astrophysical 
observations~\cite{Boddy:2014qxa}.

\section{Properties of $\mathcal{N} = 1$ SYM 
with $N_c$ colors}
\label{sec:properties}

Let us consider $\mathcal{N} = 1$ supersymmetric Yang-Mills theory 
with a gauge group SU($N_c$), where $\mathcal{N} = 1$ is the number of SUSY 
generators and $N_c$ is the number of colours.

\begin{wrapfigure}{r}{0.5\textwidth}
  \begin{center}
  \psfrag{tag1xxxxxxxxxxxxxxxxxxxx}{\footnotesize{gluino, $\kappa_\text{c}=0.13860(9)$}}
  \psfrag{tag2xxxxxxxxxxxxxxxxxxxx}{\footnotesize{a-pion, $\kappa_\text{c}=0.139264(7)$}}
  \psfrag{tag3}{\footnotesize{Vol=$16^3\times 32$, $\beta=4.0$}}
    \includegraphics[width=7.5cm]{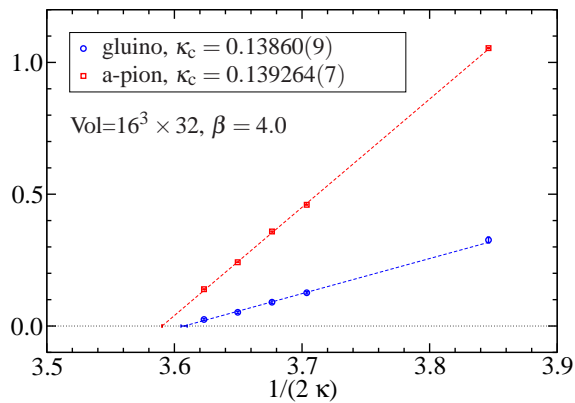}
  \end{center}
  \caption{Critical value of $\kappa_\text{c}$ defined as the value of $\kappa$
    where the square of the adjoint pion mass $m_\api^2$ (red square) and 
    the quantity $a m_S Z_S^{-1}$, proportional to the renormalised value of the 
    gluino mass, vanish (blue circle), for $N_c=3$ colours.}
  \label{fig:WTI}
\end{wrapfigure}

It is characterised by two fields: one describing the gluon $A_\mu(x)$, 
the other one describing its superpartner, the gluino $\lambda(x)$. 
The latter particle is a Majorana fermion in the adjoint representation. 
For non-zero fermion mass, SUSY is softly broken. 
Because there is only one Majorana flavour, the global chiral symmetry 
is simply U(1)$_\lambda$. It is possible to show that this symmetry is anomalous
and that, however, a $Z_{2 N_c}$ subgroup of U(1)$_\lambda$ is unbroken.
As conjectured in~\cite{Witten:1982df} the $Z_{2 N_c}$  symmetry is spontaneously
broken down to $Z_2$. The result is a theory with $N_c$ vacua, where the 
gluino condensate can be used as an order parameter (using Weyl notation):
\beq
\langle \lambda^\alpha \lambda_\alpha \rangle = \text{const} \ \Lambda^3 
e^{\frac{2 \pi \ \I}{N_c}k} \ , 
\label{eq:condensate}
\eeq
where $k=0, \dots, N_c-1$, and $\Lambda$ is a scale parameter similar to 
the $\Lambda$ parameter in QCD. 

This result is supported by arguments in different approaches, see
{\it e.g.} \cite{Shifman:1987ia, Davies:1999uw}.
The coexistence of these $N_c$ vacua implies a 
first order phase transition (for a massless gluino).
We have confirmed this behaviour in the SU(2) 
theory~\cite{Kirchner:1998mp}
and we had some evidence~\cite{Feo:1999hw} also for the SU(3) theory.

Another feature conjectured for these theories is confinement: the particle spectrum of the theory consists of colourless bound states. In the SUSY limit, the particles are organised in mass-degenerate multiplets.
In~\cite{Veneziano:1982ah} the authors wrote down an effective 
action and derived a first supermultiplet of the low-lying spectrum. 
It consists of a scalar ($0^+$ gluinoball: $\afn \sim \bar\lambda \lambda$),
a pseudoscalar ($0^-$ gluinoball: $\aetap \sim \bar\lambda \gamma_5 \lambda$),
and a Majorana fermion (spin $1/2$ gluino-glueball:
$\chi \sim \sigma^{\mu \nu} \tr{ F_{\mu \nu} \lambda }  $).
A second supermultiplet was introduced in~\cite{Farrar:1997fn} based on 
pure gluonic states in the effective action. It consists of
a $0^-$ glueball, a $0^+$ glueball, and again a gluino-glueball.
According to the authors, this last multiplet should be lighter than the 
previous one;
other authors~\cite{Feo:2004mr}, using different arguments, and hints from 
ordinary QCD, deduce the opposite order: 
clarifying this issue is one of the tasks of our project.

\section{The theory on the lattice and simulations}

The idea that it is possible to study supersymmetric gauge theories on the 
lattice goes back to~\cite{Curci:1986sm}. The proposal was that, instead 
of trying to have some remnant of SUSY on the lattice, one should only require
that SUSY is recovered in the continuum limit, in the same way as it happens
for chiral symmetry. 
The formulation we employ in our simulations is an improved version of what 
was first proposed in~\cite{Curci:1986sm}: the gauge fields are described by 
the Wilson action but with a tree-level Symanzik improvement; the gluinos
are described by Wilson fermions in the adjoint representation.
To reduce the lattice artifacts we apply one or three levels of stout smearing 
to the link fields in the Wilson-Dirac operator.
The configurations have been obtained mainly by a two-step polynomial hybrid 
Monte Carlo (TS-PHMC) algorithm~\cite{Montvay:2005tj, Demmouche:2010sf}. 
Some results, on $6^4$ lattices, have been obtained with a 
Rational Hybrid Monte Carlo (RHMC) algorithm. 
\begin{wrapfigure}{r}{0.5\textwidth}
\vspace{-10mm}
\phantom{.}
  \begin{center}
    \includegraphics[width=7.0cm]{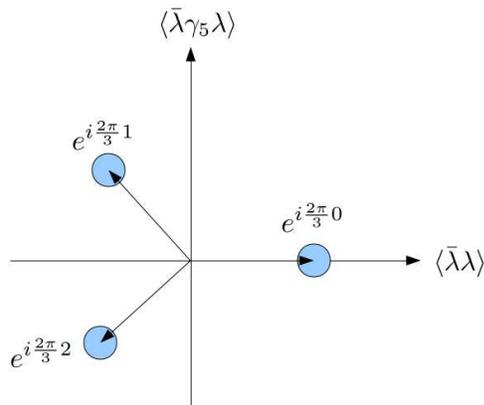}
  \end{center}
  \caption{Scatter plot expected for $\kappa=\kappa_c$.}
  \label{fig:scatterplot}
\end{wrapfigure}
Integrating out the Majorana fermions yields a Pfaffian. It can have
a negative sign, in particular for small gluino masses near $\kappa_c$,
but the significance of the negative contributions is
reduced towards the continuum limit.
If necessary, the sign is taken into account by reweighting.
Part of the work was done including a clover term: we have already verified
a great improvement in the case of SU(2)~\cite{Bergner:2015adz}.
We have simulated the theory mainly on a $16^3\times 32$ volume, with two 
values of the inverse gauge coupling $\beta$, 4.0 and 4.3, and different
values of the hopping parameter $\kappa$.

On the lattice we can identify two sources of explicitly SUSY breaking:
the first one is due to the introduction of a non-zero gluino mass;
the second one is a consequence of the breaking of the translational 
invariance for non-zero lattice spacing.
Moreover we have verified that finite volume effects can drastically 
increase the mass splitting~\cite{Bergner:2012rv}; 
in the case of SU(2), using a box size of about 1.2 fm (in QCD
units) or larger, the effect is negligible. Note that periodic boundary 
conditions in spatial directions are compatible with SUSY~\cite{Witten:1982df}. 
It is an interplay between having a finite lattice spacing 
and a finite volume that leads to the measured larger mass splitting.

\section{Tuning towards the SUSY limit}
\label{sec:tuning}

As discussed in~\cite{Curci:1986sm}, and clarified in~\cite{Suzuki:2012pc},
the chiral symmetric limit implies the supersymmetric limit. 
More precisely, both the U(1)$_\lambda$  Ward-Takahashi identity (WTI) 
(with the axial anomaly) and the SUSY WTI are restored by a 
single fine-tuning of the bare gluino mass. This gives
a solid theoretical basis for lattice formulations of $\mathcal{N} = 1$
SYM theories.

\begin{wrapfigure}{r}{0.5\textwidth}
  \begin{center}
    \includegraphics[width=7.5cm]{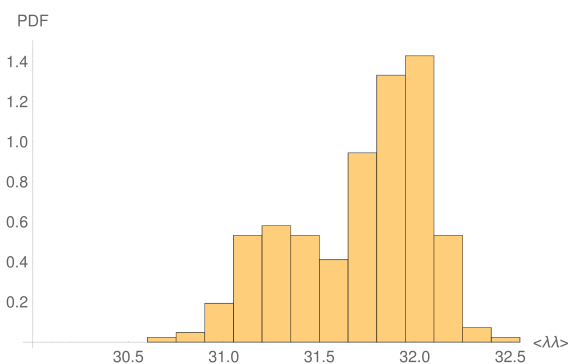}
  \end{center}
  \caption{Scalar condensate distribution for $\kappa \lesssim \kappa_c$ 
obtained using the RHMC algorithm
with $\beta=5.6$, $\kappa=0.1658$ and $c_{sw}=1.587$ on a $6^4$ volume.}
  \label{fig:distrC}
\end{wrapfigure}

In practice, one has to tune the bare gluino mass so that 
the renormalised gluino mass vanishes. As discussed in~\cite{Veneziano:1982ah}
using the OZI approximation, or in~\cite{Munster:2014cja} using a 
partially quenched setup, the square of the adjoint pion mass $m_\api^2$ 
is proportional to the mass of the gluino. 
This relation is apparently well satisfied, as shown in Figure~\ref{fig:WTI} 
(the red dashed line is a fit to the five data points),  
where $m_\api^2$  is linear in $1/(2\kappa)$. 
Given the previous relation, we can determine the critical hopping parameter
$\kappa_c$, where the renormalised mass of the gluino vanishes.
A more direct tuning approach is, of course, to determine
the renormalised gluino mass using the WTI. The procedure has already been
described in~\cite{Farchioni:2001wx}; in Figure~\ref{fig:WTI} we show directly
the result: the quantity $a m_S Z_S^{-1}$ is plotted against $1/(2\kappa)$.
$m_S$ is called subtracted mass and can be identified with the renormalised
mass of the gluino; $Z_S$ is a multiplicative renormalisation coefficient.
We see a clear linear dependence between $a m_S Z_S^{-1}$ and $1/(2\kappa)$:
fitting the data (the blue dashed line) we can determine the critical
value of $\kappa$. 
Comparing the results obtained using the two methods we see that they are 
compatible only in $7.5 \sigma$. Terms in the WTI proportional to the 
lattice spacing of course contribute to the difference between the
results and there might be also some finite size effects. In future work
a better control of systematic errors is needed.
Because the determination of the adjoint pion mass is much simpler
and numerically not expensive, it will be used as our standard method 
to tune the SUSY limit in this theory.

\begin{figure}[ht]
\centering
\includegraphics[width=5.2cm, angle=0]{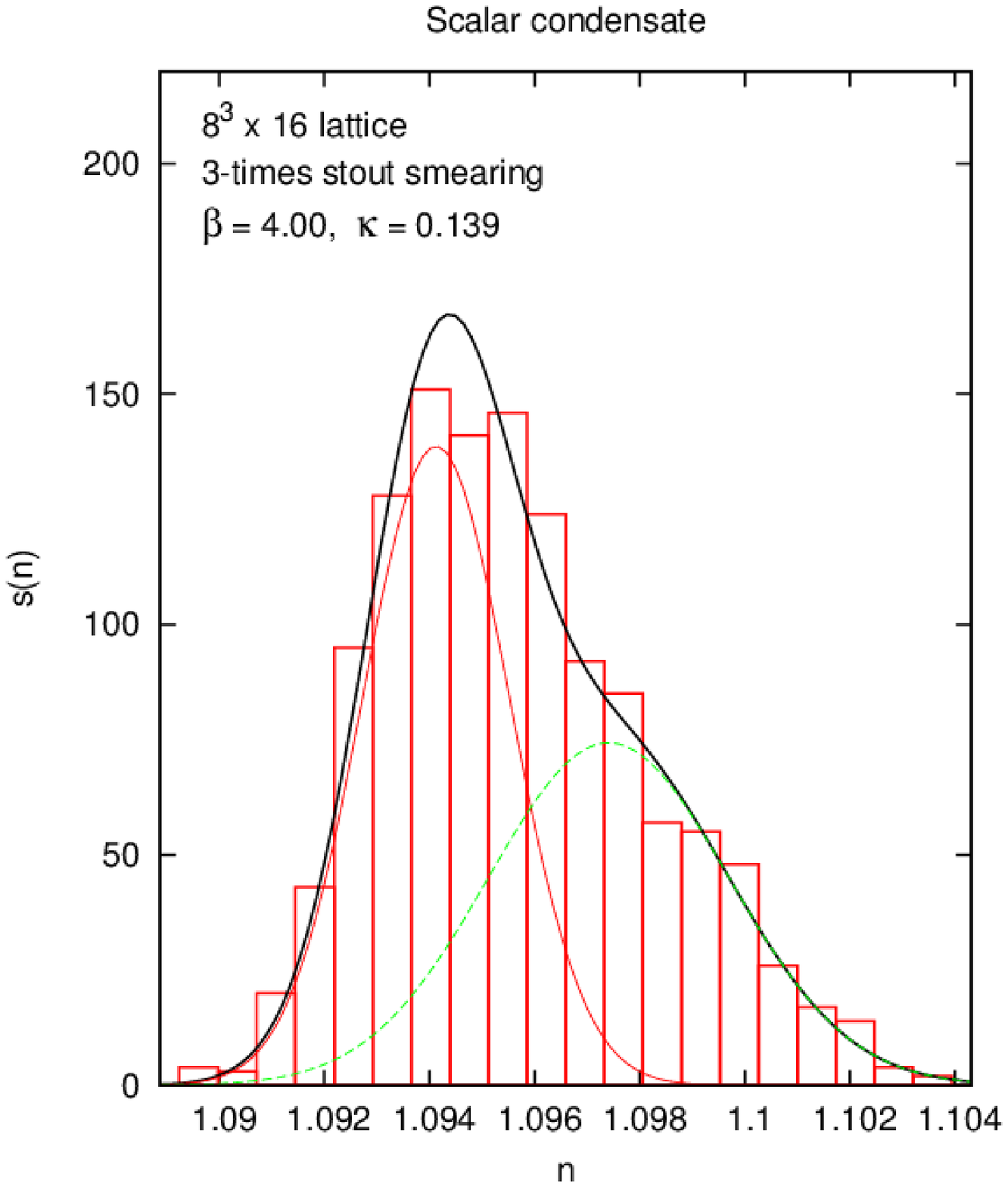}
\hspace{10mm}
\includegraphics[width=5.0cm, angle=0]{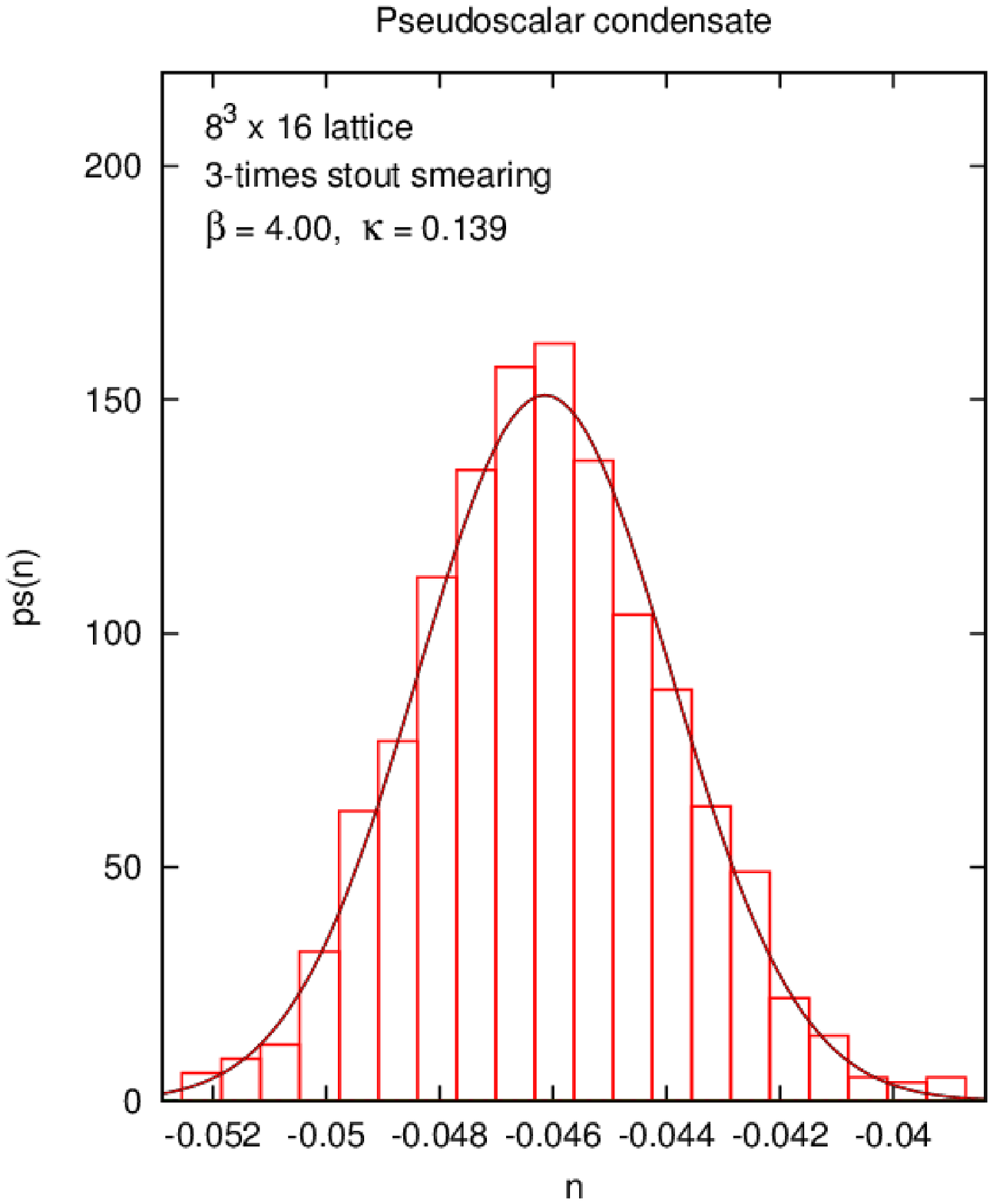}
\caption{\textbf{(Left)} Scalar condensate distribution: this case corresponds
to $\kappa \simeq \kappa_c$ where the peak on the left (in black) is about
two times higher than the one on the right (in green).
\textbf{(Right)} Pseudoscalar condensate distribution: we see only one peak, 
while two smaller peaks at the two sides should appear.}
\label{fig:distrAB}
\end{figure}

\section{The vacuum of the theory}

Eq.~\ref{eq:condensate} when translated into the Dirac 
representation, gives rise to two distinct condensates: a scalar condensate
$\langle \bar\psi\psi \rangle$ and a pseudoscalar condensate 
$\langle \bar\psi \gamma_5 \psi \rangle$.
As discussed in Sec.~\ref{sec:properties}, we expect that our 
$\mathcal{N} = 1$ SYM theory with gauge group SU(3) is characterised 
by $N_c=3$ vacua.
The three vacua lie in a Cartesian plane, according to Eq.~\ref{eq:condensate}, 
where we put on the abscissa the scalar condensate and on the ordinate
the pseudoscalar condensate, see Figure~\ref{fig:scatterplot}.

A first order phase transition should show up as a jump in the
expectation value of the scalar gluino condensate at $\kappa=\kappa_c$. Looking
at the distribution of this quantity, in relatively small volumes, 
one expects to see a two peak structure.
This is possible only in a small volume; increasing it, the
tunnelling between the three ground states becomes less probable and 
above a certain value practically impossible.

For $\kappa < \kappa_c$ we expect $\langle \bar\psi\psi \rangle > 0$ and
only one peak should appear (corresponding to the distribution labelled 
with $e^{\I \frac{2 \pi}{3} 0 }$ in Figure~\ref{fig:scatterplot} ). 
When $\kappa \rightarrow \kappa_c$ a second peak should emerge 
(corresponding to the sum of the distributions labelled 
with $e^{\I \frac{2 \pi}{3} 1 }$ and $e^{\I \frac{2 \pi}{3} 2 }$ 
in Figure~\ref{fig:scatterplot}).
This is exactly what appears in Figure~\ref{fig:distrC}. This result
was obtained on a small volume $6^4$ using our RHMC algorithm.
For $\kappa = \kappa_c$ we should see two peaks, with the left one two 
times higher the right one: this can be seen in 
Figure~\ref{fig:distrAB} (Left) where data had to be fitted as a sum 
of two Gaussian distributions.

When we look at $\langle \bar\psi \gamma_5 \psi \rangle$ we expect 
to see only one peak for $\kappa < \kappa_c$
(corresponding to the distribution labelled 
with $e^{\I \frac{2 \pi}{3} 0 }$ in Figure~\ref{fig:scatterplot}), 
but for  $\kappa \rightarrow \kappa_c$ three peaks should appear:
one in the center and two, with the same height,
at the two sides (corresponding to the two distributions labelled  
with $e^{\I \frac{2 \pi}{3} 1 }$ and $e^{\I \frac{2 \pi}{3} 2 }$ 
in Figure~\ref{fig:scatterplot} ).
The reason because they have the same height is simple:
we are populating two vacua which have the same probability
to be occupied. 
So far we have never observed the double peak structure, but only
one symmetric peak, as in Figure~\ref{fig:distrAB} (Right). 
This happen even for $\kappa > \kappa_c$ where the peak, corresponding to 
the distribution labelled with $e^{\I \frac{2 \pi}{3} 0 }$ in 
Figure~\ref{fig:scatterplot}, should disappear and only the two peaks
at its sides should remain. 
The reason for this phenomenon is still under investigation.

\section{Mass spectrum: numerical results}

As discussed in Sec.~\ref{sec:tuning} the 
renormalised mass of the gluino is proportional to the square of the 
adjoint pion mass. 
As a consequence we extrapolate the masses of the bound states
to the chiral limit fitting their mass against the square of the 
adjoint pion. We present here only the results obtained for $\beta=4.0$.

\begin{figure}[ht]
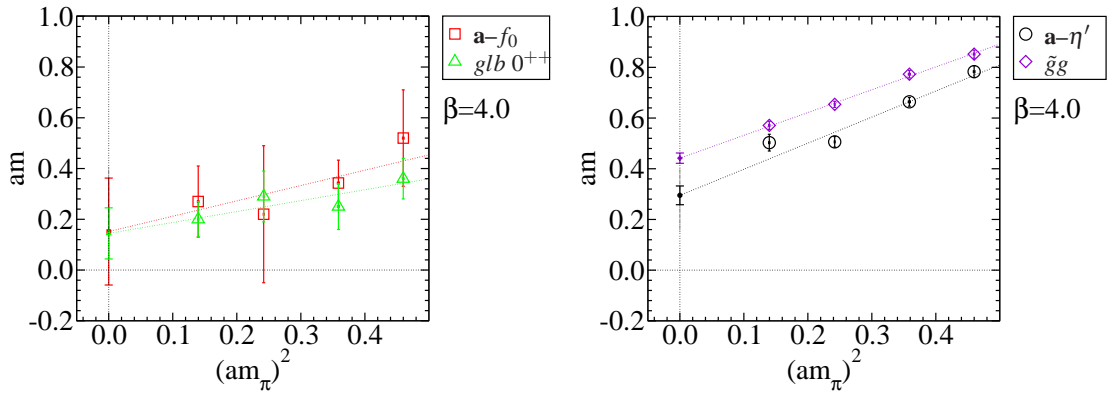

\centering
\psfrag{GTG}{\footnotesize{\textbf{$\tilde{g}g$}}}
\psfrag{GGPP}{\footnotesize{\textbf{$glb \ 0^{++}$}}}
\psfrag{ETA}{\footnotesize{\textbf{$\aetap$}}}
\psfrag{EF0}{\footnotesize{\textbf{$\afn$}}}
\includegraphics[width=7.2cm, angle=0]{./figures/chiral_f0_zpp_190b.eps}
\hspace{2mm}
\includegraphics[width=7.0cm, angle=0]{./figures/chiral_gg_eta_190b.eps}
\phantom{.}
\caption{\textbf{(Left)} Mass of the $\afn$ and of the glueball $0^{++}$
for different values of the square of the adjoint pion mass $m_\api$.
 \textbf{(Right)} As for (Left) but for the mass of the $\aetap$ 
and of the gluino-glueball $\tilde{g}g$.}
\label{fig:spectrum}
\end{figure}

The mass spectrum of this theory has been determined using the techniques
already discussed in~\cite{Bergner:2015adz}. 
In Figure~\ref{fig:spectrum} (Left) we plot the channel $0^{++}$, showing
both the glueball and the $\afn$, for different values of $m^2_\api$ . 
It is well known that this channel is particularly noisy;   
in our case the glueball determination looks somewhat better than the meson 
one. The masses of both states are perfectly compatible when extrapolated 
to the chiral limit.

In Figure~\ref{fig:spectrum} (Right) we plot the mass of the $\aetap$
and of the gluino-glueball. Their error bars are considerably smaller than
those in the channel $0^{++}$; the gluino-glueball is by far the state 
with the best results.
The masses of these two states, when extrapolated to the chiral limit, 
are roughly compatible with those in the $0^{++}$ channel, taking the errors in 
that channel into account. But their masses are not compatible with each other.
This is probably due to discretisation effects, which should be already 
reduced at $\beta=4.3$ and that we are currently simulating.

\section{Conclusions and outlook}

We have presented our first results on $\mathcal{N} = 1$ supersymmetric 
Yang-Mills theory with three colours. We have shown results on 
the use of the Ward-Takahashi identities and of the adjoint pion to tune 
the theory to supersymmetry.
The structure of vacua has been investigated and some preliminary results 
have been obtained: we see a clear first order transition in the scalar 
condensate but not in the pseudoscalar one.
This issue is still under investigations. 
We started to measure the particle spectrum of the theory. At the
moment we presented the results with only one lattice spacing but the
results are promising. 
We are currently simulating the theory with a second value of
$\beta$ to estimate the discretization effects 
and, hopefully, to extrapolate the results to the continuum limit.

\section*{Acknowledgements}
The authors gratefully acknowledge the Gauss Centre for Supercomputing (GCS) 
for providing computing time for a GCS Large-Scale Project on the GCS share 
of the supercomputer JUQUEEN at J\"ulich Supercomputing Centre (JSC) and on 
the supercomputer SuperMUC at Leibniz Computing Centre (LRZ). GCS is the 
alliance of the three national supercomputing centres HLRS (Universit\"at 
Stuttgart), JSC (Forschungszentrum J\"ulich), and LRZ (Bayerische Akademie 
der Wissenschaften), funded by the German Federal Ministry of Education and 
Research (BMBF) and the German State Ministries for Research of 
Baden-W\"urttemberg (MWK), Bayern (StMWFK) and Nordrhein-Westfalen (MIWF).
Further computing time has been provided by the computer cluster PALMA of 
the University of M\"unster.

\end{document}